\documentclass[aps,prb,twocolumn,showpacs]{revtex4}

\usepackage{amsmath}
\usepackage{amssymb}
\usepackage{bm}
\usepackage{epsfig}
\usepackage{graphicx}
\usepackage{color}

\usepackage{color}

\def\be{\begin{equation}}
\def\ee{\end{equation}}
\def\bea{\begin{eqnarray}}
\def\eea{\end{eqnarray}}

\begin{document}
\newcount\timehh  \newcount\timemm
\timehh=\time \divide\timehh by 60
\timemm=\time
\count255=\timehh\multiply\count255 by -60 \advance\timemm by \count255

\title{Ground state of the holes localized in II-VI quantum dots \\ with  Gaussian potential profiles}
\author{M. A. Semina, A.A. Golovatenko and A. V. Rodina}
\affiliation{Ioffe Institute, 194021, St.-Petersburg, Russia. }

\begin{abstract}
We report on the theoretical study of   the hole states in  II-IV quantum dots
of a spherical and ellipsoidal shape, described by a smooth  potential confinement profiles, that can be modelled  by a
Gaussian functions in all three dimensions. The 
universal dependencies of the hole energy, $g$-factor and localization length  on a
quantum dot barrier height, as  well as the ratio of effective masses of the light and heavy holes  are presented for the spherical quantum dots. The splitting of the four-fold degenerate
ground state into two doublets is derived for anisotropic (oblate or prolate) quantum dots. Variational calculations are combined with numerical ones 
 in the framework of the Luttinger Hamiltonian. Constructed trial functions
are optimized by comparison with the numerical results. The effective hole $g$--factor is found to be independent on the quantum dot
size and barrier height and is approximated by simple universal expression depending
only on the effective mass parameters. The results  can be used for
interpreting and analyzing  experimental spectra measured in various
structures with the quantum dots of different semiconductor materials.
\end{abstract}


\maketitle

\section{Introduction}

Quantum dots (QDs), sometimes referred to  as nanocrystals (NCs) in literature, are systems with good prospects for nanotechnology. Physically, the QDs are tiny semiconductor nanoparticles which are formed in various dielectric or semiconductor matrices by different methods. Among them, the basic ones are the chemical synthesis and  epitaxy used to fabricate, respectively, the colloidal and epitaxial QDs. Depending on the used semiconductors and the fabrication methods, the QDs can have various shapes and sizes. Carrier localization in the QDs makes interaction between charge carriers more efficient as compared with bulk materials, and  many effects that are weak in bulk materials become observable. Thus, the development of new methods of the calculation of the wave functions and energy spectra of charge carriers is very important for designing the QD structures.

 The incentive to our study was the renewed interest in II-VI QDs for various applications. In particular, the epitaxial CdSe/Zn(S,Se) QDs have been successfully used as an active region in laser heterostructures pumped optically or by an electron beam. \cite{Ivanov2013,Sorokin2015} Besides, they have been recognized as promising candidates for room temperature single photon emission and production of photon pairs due to strong carrier confinement and distinct biexciton performance \cite{Strauf,Tribu,Fedorych}. The self-formation of these nanostructures takes place when a CdSe insertion of a fraction monolayer (ML) thickness is deposited within a Zn(S,Se) matrix. Previously, this insertion was considered as a disordered quantum well, where the nano-islands with high Cd content, $x$, are formed within the matrix with lower $x$. \cite{Klochikhin,Peranio2000} Thorough transmission electron microscopy (TEM) studies \cite{Litvinov2008}, however, have shown that the Cd content in the Cd-rich nano-islands can be as high as 80\% in the center, decreasing towards the periphery, while the Cd concentration in the surrounding area approaches 20\%
 only. The average lateral sizes of the nano-islands are about 5 nm, while the sizes in the growth direction are somewhat smaller. These findings make it possible to consider the nano-islands as oblate ellipsoidal QDs. Such an asymmetric shape influences the energy splitting of both exciton and biexciton states in the single CdSe/ZnSe QD \cite{Kulakovskii99}, that is important for the generation of the entangled photon pairs. The CdSe/Zn(S,Se) structures exhibited the long-lived electron spin coherence, related likely to the three-dimensional localization potential \cite{Syperek}. Recently, it has been demonstrated that at  a certain deposited amount of CdSe the nano-islands can be quite isolated \cite{Reznitsky2015}. Importantly, the change of the concentration between the dot and barrier regions in the epitaxial CdSe/Zn(S,Se) QDs is not abrupt but gradual due to diffusion and segregation processes.

The II-VI QDs  with such a gradual composition variation are expected to
demonstrate an improved radiative emission. Indeed, for a long time, the
biexciton performance   of the chemically synthesized   colloidal QDs was suffering from
a high rate of non-radiative Auger recombination. To overcome this problem colloidal
CdSe-based nanocrystal heterostructures  with gradually changing composition were
synthesized and reported in Ref. \onlinecite{Nasilowski15}. The non-radiative
Auger processes are suppressed in such structures due to smoothing of the confining
potential. \cite{Cragg2010,Garcia-Santamaria2011,Bae,Climente2012} Further progress
in analysis of optical phenomena in the II-VI QDs and manufacturing the efficient nano
emitters of quantum light requires the elaborated model description of quantum states in
the ellipsoidal QDs with smooth potential profile.
Among variety of theoretical methods, including atomistic tight-binding
\cite{Korkusinski10,Zieli?nski15} or pseudo-potential calculations \cite{Bester10}, and
$\bm{kp}$-theory, the latter provides a reasonable compromise between the accuracy and
computational complexity. The $\bm{kp}$ method is particularly suitable for
nanostructures with  a smooth potential profile, where interface effects play a minor
role.

In the most simple effective mass model of non-degenerate parabolic band,
various kinds of confinement potentials for QDs were theoretically studied:
abrupt potential with infinite and finite barriers, \cite{Efros82, Brus, Marin,
Pellegrini,Schooss94,Rego97} parabolic potential \cite{Que,Xie2006} and various
kinds of smooth potential with finite height. \cite{Adamowski, Ciurla,
Xie2009,Szafran2001} The parabolic potential was also shown to be  a good approximation for
the in-plane smooth profile  of the lens-shaped self-assembled QDs.\cite{Wojs1996}
However, to describe properly the energy spectra of the holes confined in QDs,
one has to take into account the complex structure of the valence band. In widely used semiconductors (including II-VI), the top of the
valence band is four-fold degenerate and has $\Gamma_8$ symmetry and can be described by
the Luttinger Hamiltonian. \cite{Luttinger} The fine energy structure of the hole states
defines the selection rules for inter-band transitions. Moreover,  the Zeeman splitting
in the external magnetic field is determined by the light and heavy holes splitting and
mixing. \cite{Rego97, Kubisa11,Semina2015} Therefore, the understanding of the
characteristic properties of the hole states in QDs is important for designing the
structures with required optical properties.

For  the spherical NCs \cite{Efros89,Xia89,EkimovJOSA93}  and the
disk-like QDs \cite{Rego97} modelled by abrupt potential with infinite barrier,   it was
shown, that the mixing of the heavy and light hole states can significantly modify the
energy spectrum and the wave function of the hole ground state, as well as its splitting
which can be caused by the anisotropy of QD shape, intrinsic crystal field, and applied
magnetic field.\cite{EfrosPRB96} Later, the full multiband ${\bm k}\cdot {\bm
p}$ modelling was developed for spherical NCs\cite{EfrosPRB98} and NC heterostructures
with  abrupt potential barriers,\cite{Jaskolski98} as well as for the pyramidal and
disk-shaped epitaxial QDs.\cite{Pryor98,Stier99,Tadic2002} The models of parabolic
potential with valence band degeneration taken into account were used for  QDs of
different shapes, \cite{Rinaldi,Rodina2010,Semina2015} not only within effective mass
approach but also as the modelling tool along with more elaborated atomistic
calculations.\cite{Korkusinski2013}  However, to the best of our knowledge, the
QDs with smooth but finite height potential confinement in all three spatial dimensions
have been considered  only within single band effective mass approximation so far. This simplified approximation can hardly be used for both epitaxial CdSe/ZnSe and colloidal
CdSe/CdS QDs with gradually varying composition. \cite{Peranio2000,Litvinov2008,Nasilowski15}

In present paper we consider  QDs with a shape which is close to either spherical or ellipsoidal one and a smooth potential profile  which can be described by the Gaussians in all three spatial directions. We focus on the characteristics of the holeû localized in such potentialá
taking into account the complex valence band structure in the framework of the Luttinger
Hamiltonian. It is shown that the properties of the holes localized in a smooth potential
profiles, e.g. energy splittings caused by the shape anisotropy or external magnetic
field, might be very different from the properties of the holes localized in a box--like
QDs with abrupt potential barriers.

The paper is organized as follows: in Section \ref{simple_band}  we introduce the
regularities of our problem on the most intuitive example of the material with single
band isotropic parabolic dispersion. In Section \ref{spheric_luttinger} we move on to the
characteristic properties of the hole localized in spherically symmetric quantum dot with
the  top of valence band that can be described by the Luttinger Hamiltonian. In Sections
\ref{anis_splitting} and \ref{magnetic_field} we consider the hole ground state splitting
due to the anisotropy of the QD potential, crystal field, and external magnetic field. In
the end we summarize our results.

\section{Localization of a particle in QD with parabolic or Gaussian profile: single band approximation}\label{simple_band}

To introduce the specifics of the carriers localization in the quantum dots described by the smoothly varying spatial potential $V({\bm r})$ ($r$ is the radial coordinate) we consider the  the Schr{\"o}dinger equation $[\hat H + V({\bm r})] \Psi = E \Psi$ with the single band isotropic effective mass Hamiltonian
\begin{equation}
\hat H  = \frac{\hbar^2\hat k^2} {2m^*} \, .
\end{equation}
Here ${\bm k} = -{\mathrm{i} }{\bm \nabla}$ is the wave vector operator and $m^*=m_{\rm e(h)}$ is the electron (hole) effective mass.  We consider the cases of the spherical symmetry  $V({\bm r}) = V({ r}) $ and axial symmetry $V({\bm r}) = V(\rho,z) $ potentials, where $r^2=x^2+y^2+z^2=\rho^2+z^2$, $x,y,z$ -- are Cartesian coordinates.

\subsection{Spherically symmetric QD}
\label{simple_sp}

We start with the  spherical parabolic (harmonic oscillator) potential, that is the limiting case for the Gaussian potential
\begin{equation}
V_0(r)=\frac{\kappa }{2}r^2, \label{par}
\end{equation}
 where $\kappa $ is the spring constant. In the framework of single band effective mass approximation the exact solutions of the problem with such potential are well known. The spherical oscillator  wave functions can be easily found as:
\begin{multline}
\Psi_{nlm}({\bf r})=R_{nl}(r) Y_{lm}(\Theta) \, , \\
R_{nl}(r)= \frac{1}{L^{3/2+l}}\left[\frac{2n!}{\Gamma\left(n+l+3/2\right)}\right]^{1/2}r^l\times\\  \times \exp\left[-\frac{r^2}{2L^2} \right]L_n^{l+1/2}\left[\frac{r^2}{L^2} \right] \, ,
\label{harm}
\end{multline}
and correspond to the
 equidistant eigen energies
 \begin{equation}
 E_{N}=\hbar \omega (N+3/2) \, \quad N=2n+l=0,1,2.....
\end{equation}
Here $\omega=\sqrt{\kappa /m^*}$ and $L=\sqrt{\hbar/m^*\omega}$ are the characteristic oscillator  frequency and oscillator length, respectively, $n,l$ and $m$ are principal, orbital and magnetic quantum numbers, respectively, $Y_{lm}$ are the spherical angular harmonics \cite{Edmonds} and $L_{n}^{l+1/2}$ are the generalized Laguerre polynomials \cite{polynoms}.   The ground state energy and radial wave function  of the particle  localized in $V_0(r)$  are characterized by $n=0$, $l=0$ and  given by
 \begin{equation}
E_0= \frac{3}{2}\hbar\omega=\frac{3}{2}\frac{\hbar^2}{m^*L^2}\label{e0}
\end{equation}
and
 \begin{equation}
R_0(r)=\frac{2}{\pi^{1/4}L^{3/2}} \exp\left[-\frac{r^2}{2L^2} \right] \, .
\end{equation}

\begin{figure}[!ht]
\begin{center}
\includegraphics[width=0.75\columnwidth]{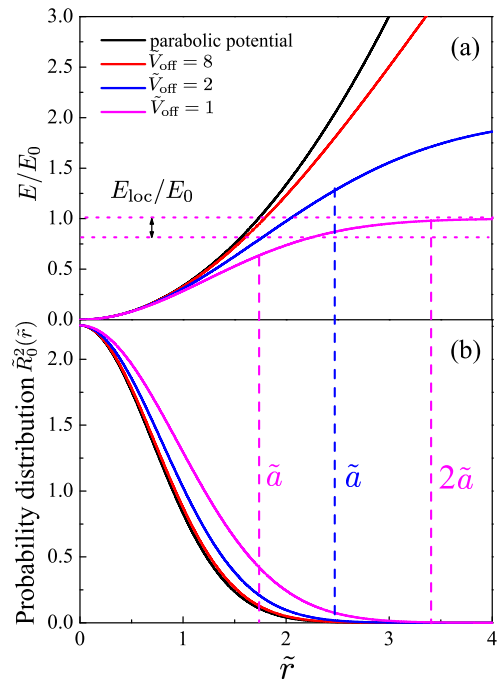}
\end{center}
\caption{\label{fig:pot1}(Color online) (a) Dimensionless  parabolic and Gaussian potentials with  $\tilde V_\text{off}=8;2;1$.  (b) Probability distribution $\tilde{R}_0^2(\tilde{r})$ for different $\tilde V_\text{off}$;   Vertical dashed lines on (a) and (b) show the characteristic radii $\tilde a$ for  $\tilde V_\text{off}=1$ and $2$ and $2 \tilde a$  for  $\tilde V_\text{off}=1$.  Horizontal dotted  lines on panel (a)  show the potential depth  $\tilde V_\text{off}$  and energy level $E/E_0$ for $\tilde V_\text{off}=1$. }\label{simple1}
\end{figure}

The parabolic potential describes the QD with smooth profile, however yet it does not permit us to consider the QDs with the finite potential barrier outside the dot. To do so we consider the potential of the
 Gaussian  form
 \begin{eqnarray}
V_\text{G}(r)= V_\text{off}\left(1-\exp\left[-\frac{r^2}{a^2} \right]\right), \label{gauss}
\end{eqnarray}
where  $V_\text{off}$ is the energy step (band-offset) between the QD center and the surrounding medium, determined  as $r > 3a$, while $a$ can be used as the rough estimation of the QD radius. Near the QD center, at $r \ll a$, the Gaussian potential can be approximated as parabolic $V_\text{G}(r) \approx V_0(r)$ with the spring constant $\kappa =2V_\text{off}/a^2$.  To simplify the comparison of the potentials characterized by the same spring constant $k$ at the QD center and different potential barriers $V_\text{off}$, we chose the parameters of the single band parabolic problem -- the ground state energy  $E_0$ and the oscillator length $L$ -- as the energy and length units for all QDs. In these units the parabolic and Gaussian potentials take the form
  \begin{eqnarray}\label{dimless}
  \tilde V_0(\tilde r)= V_0(r/L)/E_0 = \frac{1}{3} \tilde r^2 \, , \\
\label{dimless1}\tilde V_\text{G}(\tilde r)= V_\text{G}(r/L)/E_0=\tilde V_\text{off}\left(1-\exp\left[-\frac{\tilde r^2}{3\tilde V_\text{off}}\right]\right),
\end{eqnarray}
 where $\tilde V_\text{off}=V_\text{off}/E_0$ and $\tilde r=r/L$. Note that the spring constant is  not included explicitly in eqs. \eqref{dimless} and \eqref{dimless1}. It is, however, contained in expressions for our units $L$ and $E_0$. This fact allows us to obtain the universal dependence of the localized particle wave function and energy spectrum on  $\tilde{V}_{\text{off}}$.
The dimensionless parabolic and Gaussian potentials with different $\tilde V_\text{off}$ are shown in Fig. \ref{fig:pot1}(a), the chosen values of $\tilde{V}_{\text{off}}$ correspond to characteristic radii $\tilde a=a/L = \sqrt{3\tilde V_{off}}\approx 1.73,~2.45$ and $4.9$. Vertical dashed lines show the characteristic radii $\tilde a$ for shallowest dots with $\tilde V_\text{off}=1$ and $2$, for $\tilde V_\text{off}=8$ the value of $\tilde{a}$ lays outside the scale of the figure, the parabolic potential has the infinite effective radius.

\begin{figure}[!ht]
\begin{center}
\includegraphics[width=0.85\columnwidth]{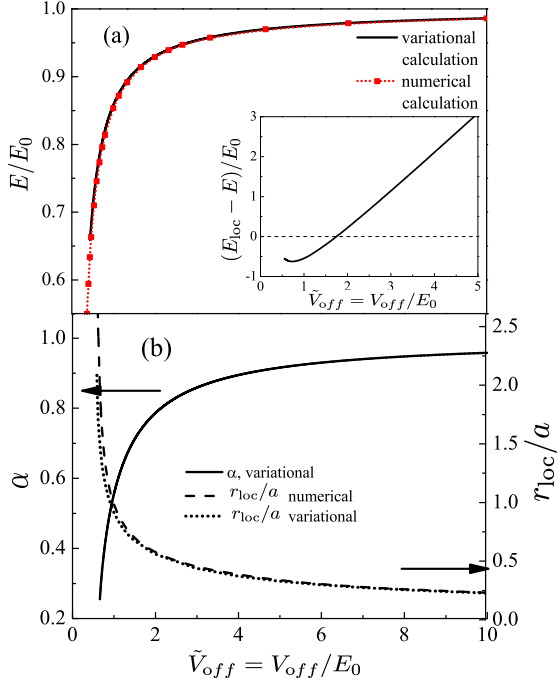}
\end{center}
\caption{\label{fig:pot2}(Color online)  (a) Dependence of the hole ground state energy $E/E_0$ on the $\tilde V_\text{off}=V_\text{off}/E_0$, the inset shows the dimensionless difference between the localization energy and particle energy; (b) Dependencies of the variational parameter $\alpha$ and the localization radius $r_{\rm loc}/a$   on $\tilde V_\text{off}$. }\label{simple2}
\end{figure}

	Since no exact solution exists for the particle in Gaussian potential, we found the wave function and  energy of the ground state by two methods - numerical and variational. Numerical solution can be found by expanding the radial wave functions  over the basis of the oscillator functions \eqref{harm} and diagonalizing the resulting matrix.
	
To obtain the ground state energy by the  variational procedure we chose the dimensionless probe function  $\tilde{R}_0(\tilde{r})=R_0(r/L)L^{3/2}$ in the form:

 \begin{equation}  \label{R0}
\tilde{R}_0=\frac{2 \alpha^{3/4}}{\pi^{1/4}} \exp\left[-\frac{\alpha \tilde{r}^2}{2} \right] \,
\end{equation}
with $\alpha$ being the only trial parameter. With the probe function \eqref{R0} we have the following expression for the particle ground state energy as a function of $\alpha$:
\begin{equation}\label{Ea}
E(\alpha)/E_0=\frac{\alpha}{2}+\tilde{V}_\text{off}-\frac{3\sqrt{3}\alpha^{3/2}\tilde{V}^{5/2}_\text{off}}{\left(1+3\alpha \tilde{V}_\text{off}\right)^{3/2}}.
\end{equation}

Figure  \ref{fig:pot1}(b) shows the dimensionless probability distribution $\tilde{R}_0^2(\tilde{r})$ for Gaussian QDs with different $\tilde V_\text{off}$ and for parabolic QDs.  Figure \ref{fig:pot2}(a) shows the dependences of the dimensionless ground state energy $E/E_0$ on the $\tilde V_\text{off}=V_\text{off}/E_0$  and  Fig. \ref{fig:pot2}(b) shows the dependence of the minimizing value of trial parameter $\alpha$ on $\tilde V_\text{off}$.   Panels (a)  of figs. \ref{fig:pot1} and \ref{fig:pot2} demonstrate, that with the increase of $\tilde{V}_\text{off}$ the ground state energy and the wave function in Gaussian potential tend to those of the harmonic oscillator. While with decrease of $\tilde{V}_\text{off}$ the quantization energy $E/E_0$ decreases, the localization energy $E_{\rm loc}/E_0 = (V_{\rm off} -E)/E_0$  also decreases and becomes smaller than  $E$ for $\tilde{V}_\text{off}<2$ (see inset in fig. \ref{fig:pot2}(a)). The bound state exists up to $\tilde{V}_\text{off}\approx 0.55$ as it is shown by the numerical calculation; the variational calculation gives the bound state up to $\tilde{V}_{\text{off}}\approx 0.65$.

The localization  of the particle inside the QD can be  characterized by the  localization  radius $\tilde r_{\rm loc}=r_{\rm loc}/L =\sqrt{<\tilde r^2>}=\sqrt{\int_0^\infty \tilde{R}_0^2 \tilde r^4 d \tilde r }$.  
We find $\tilde r_{\rm loc} \approx 1.23$ for the parabolic potential and $\tilde{V}_\text{off}=8$, and $\tilde r_{\rm loc} \approx 1.39$ and $\tilde r_{\rm loc} \approx 1.76$ for $\tilde{V}_\text{off}=2$ and  $\tilde{V}_\text{off}=1$, respectively. One can see that $\tilde r_{\rm loc} \approx 1.76$ only slightly exceeds the effective radius $\tilde a \approx 1.73$ for $\tilde{V}_\text{off}=1$ and $\tilde r_{\rm loc} < \tilde a $ for $\tilde{V}_\text{off}=2$ and higher potential barriers. In spite of the very small localization energy in the dots with $\tilde{V}_\text{off}<2$,  the probability density $|\tilde{R}_0(\tilde{r})|^2$ is well localized inside these dots up to $\tilde{V}_\text{off}=1$. The  dependence of $r_{\rm loc}/a=\tilde r_{\rm loc}/\tilde a$ on  $\tilde{V}_\text{off}$ is also shown in Fig. \ref{fig:pot2}(b). Note that the radius $2a$ corresponds to the point where the Gaussian potential saturates: $V_{\rm G}(2a) \approx V_{\rm off}$. For the probe function \eqref{R0}, $\tilde r_{\rm loc}=\sqrt{1.5/\alpha}$ and $r_{\rm loc}/a=\tilde r_{\rm loc}/\tilde a=\sqrt{1/(2\alpha\tilde{V}_\text{off})}$.

\subsection{Axially symmetric non-spherical QD}

Now we consider ellipsoidal axially symmetric QDs, where  parabolic confinement potential can be written as:
 \begin{eqnarray} \label{an}
V_p^a(r,z,\mu)=\frac{\kappa _{\rho}}{2}\rho^2+\frac{\kappa _{z}}{2}z^2= V_0(r)+\Delta V_p^a(\rho,z,\mu) \, , \\ \Delta V_p^a(\rho,z,\mu) =  \kappa  \mu \left(z^2- \frac{1}{3}r^2 \right).
\end{eqnarray}
Here the average spring constant is $\kappa =(2\kappa _\rho+\kappa _z)/3$ and we introduce the QD anisotropy parameter as
 \begin{eqnarray}\label{k}
\mu=\frac{3}{2} \frac{(\kappa _z/\kappa _\rho-1)}{(\kappa _z/\kappa_{\rho}+2)} =  \frac{(\kappa _z-\kappa _\rho )}{2\kappa }.
\end{eqnarray}
One can see that $\mu >0$ corresponds to the case $\kappa _z>\kappa _\rho$ and thus describes stronger confinement along $z$ direction  (oblate QD). In the opposite case of $\mu<0$ the confinement is stronger in $xy$ plane (prolate QD). Note that the Eq. \eqref{an} is exact.  It follows from  \eqref{k} that
 \begin{eqnarray}\label{kzkr}
\kappa _{\rho}=\kappa \left(1-\frac{2}{3}\mu\right),~  \kappa _{z}=\kappa \left(1+\frac{4}{3}\mu\right) \,
\end{eqnarray}
and the condition that $k_{\rho,z}\geq 0$ leads to only $-3/4\leq \mu \leq 3/2$ having physical sense.

The exact solutions for the axially symmetric harmonic potential \eqref{an} are also well known. We introduce parameters $L_x,~L_y, ~L_z$ as oscillator lengths along $x,~y,~z$ axes correspondingly, and note that in the axially-symmetric potential $L_x=L_y=L$.  In this case, the wave functions can be  calculated from the Schr{\"o}dinger equation for axially-symmetric harmonic oscillator:
\begin{multline}\label{eigenfunctions_a}
\Psi_{n_x,n_y,n_z}({x,y,z})=\frac{1}{\sqrt{2^{n_x+n_y+n_z}}n_x!n_y!n_z!}\frac{\pi^{-3/4}}{L\sqrt{L_z}}\times\\ \times H_{n_x}\left[\frac{x}{L} \right]H_{n_y}\left[\frac{y}{L} \right]H_{n_z}\left[\frac{z}{L_z} \right]\exp\left[-\frac{x^2+y^2}{2L^2}-\frac{z^2}{2L_z^2}\right] \, ,
\end{multline}
and correspond to the
 equidistant eigen energies
 \begin{multline}
 E_{n_x,n_y,n_z}=\hbar \omega (n_x+n_y+2)+ \hbar \omega_z (n_z+1), \\ \quad n_x,n_y,n_z=0,1,2.....
\end{multline}
Here $H_{n}\left[x\right]$ are Hermite polynomials \cite{polynoms}, $\omega=\sqrt{k_\rho/m}$, $\omega_{z}=\sqrt{k_z/m}$ and $L=\sqrt{\hbar/m\omega},~L_{z}=\sqrt{\hbar/m\omega_{z}}$. For the ground state energy we obtain:
 \begin{multline}\label{Emu}
E_0^a=\frac{1}{2} \left[ \frac{2\hbar^2}{mL^2} + \frac{\hbar^2}{mL_z^2} \right]= \\= \frac{2E_0}{3} \left[ \sqrt{1-2\mu/3} + \frac{1}{2} \sqrt{1+4\mu/3} \right]\approx E_0\left[1-\frac{\mu^2}{9}  \right]\, .
\end{multline}
One can see that \eqref{Emu} containes no linear on $\mu$ correction to the ground state energy. The same result can be readily observed  by treating $\Delta V_p^a(\rho,z,\mu)$ as perturbation.

We consider the anisotropic  Gaussian potential with axial symmetry in the form
\begin{eqnarray}\label{anGfull}
V_G^a(r,z,\mu)=V_{\text{off}}\left(1-\exp\left[-\frac{x^2+y^2}{a_x^2}-\frac{z^2}{a_z^2}\right]\right) \, .
\end{eqnarray}
The ground state energy of the particle in such potential can be found numerically by expanding the wave functions  over the basis of the oscillator functions \eqref{eigenfunctions_a}, diagonalizing the resulting matrix. The anisotropy can also be considered in the framework of the perturbation theory by two ways. One way is to find  the isotropic and anisotropic part of the Hamiltonian Eq. (\ref{anGfull})  as it was done in \eqref{an}:
\begin{multline}\label{anG}
V_G^a(r,z,\mu)=V_{\text{off}}\left(1-\exp\left[-\frac{x^2+y^2}{a_x^2}-\frac{z^2}{a_z^2}\right]\right)\approx \\ \approx  V_G(r)+\Delta V_p^a(r,z,\mu) \, , \\ \Delta V_G^a(r,z,\mu) = \exp\left[-\frac{r^2}{a^2}\right] \Delta V_p^a(r,z,\mu)  \, ,
\end{multline}
 where $a={\sqrt{3}a_xa_z}/{\sqrt{a_x^2+2a_z^2}}$. The effective spring constants are introduced by analogy with the spherical QD: $\kappa_{\rho}= 2V_{\text{off}}/a_x^2$ and $\kappa_{z}= 2V_{\text{off}}/a_z^2$. Anisotropy parameter $\mu$ is defined in the same way as for the parabolic potential \eqref{k}.
Approximate expansion \eqref{anG} keeps only terms linear on $\mu$ and is applicable for $\mu<1$. Again,  the  first order energy correction to the $s$ symmetry ground state for the perturbation $\Delta V_G^a(r,z,\mu)$ vanishes. Using numerical approach   for  $\mu<1$ we found that  the shift of the ground state energy in the anisotropic Gaussian potential can be described  as $E^a(\mu) \approx E^a(\mu=0)\left[1-{\mu^2}/{9}  \right]$ by analogy with expression \eqref{Emu}.

Alternatively, the anisotropy of the Gaussian potential can be treated by replacing the coordinates as $x\longrightarrow x (a_x/a)$ and $z\longrightarrow z (a_z/a)$. The potential energy becomes isotropic in the new coordinates. However, the kinetic energy operator $\hat H$ in acquires  the additional term
\begin{equation}
\Delta \hat H^{\rm a}_{\rm k}  = \frac{2\mu}{3} \frac{\hbar^2 } {2m} \left[ \hat k^2 - 3 \hat k_z^2 \right] \, . \label{axis}
\end{equation}
Again, the linear on $\mu$ energy correction to the  ground state  from $\Delta \hat H^{\rm a}$  vanishes.

\section{Spherical symmetry problem for the $\Gamma_8$ valence band.}\label{spheric_luttinger}

We consider now the hole  in four-fold degenerate $\Gamma_8$ valence subband for semiconductors with large spin-orbit splitting.
 Luttinger Hamiltonian for  such semiconductors in spherical approximation can be written \cite{Luttinger,Gelmont71} as
\begin{equation}\label{luttinger}
\hat H_{\rm L}=\frac{\hbar^2}{2m_0}\left[\left(\gamma_1 + \frac{5}{2}\gamma\right) \hat k^2 -2\gamma (\hat {\bm k}{\bm j})^2 \right]  \, .
\end{equation}
Here $m_0$ is the free electron mass, ${\bm j}$ is the hole internal angular momentum operator for $j=3/2$,  $\gamma_1$ and $\gamma=(2\gamma_2+3\gamma_3)/5$ are Luttinger parameters related to the light and heavy hole effective masses as $m_{lh,hh} = m_0/(\gamma_1 \pm 2 \gamma)$.

	 The first energy level of holes in a spherical QDs in a semiconductor with degenerate $\Gamma_8$ valence subband is $1S_{3/2}$ state.\cite{Efros89,EkimovJOSA93} It has total angular momentum ${\bm J}={\bm j}+{\bm l}$ with $J=3/2$ and is four-fold degenerate with respect to its projection  on the $z$ axis. The wave  functions of this state can be written as\cite{Gelmont71,note}
\begin{multline}
\Psi_{M} = 2 \sum_{l=0,2} (-1)^{M-3/2} (i)^l R_l(r)\times \\ \times \sum_{m+\mu = M}
\left(
\begin{array}{ccc}
l & 3/2&3/2 \\ m&\mu&-M
\end{array}
\right)
Y_{l,m} u_\mu \, ,
\label{holewf}
\end{multline}
where   $\left(_{m~n~p}^{i~~k~~l}\right)$ are the Wigner 3j-symbols, and $u_\mu $
($\mu = \pm 1/2, \pm 3/2$) are the Bloch functions of the four-fold
degenerate valence band $\Gamma_8$ that can be found in Ref. \onlinecite{Ivchenko}.
The radial wave functions $R_0$ and $R_2$ in Eq. \eqref{holewf} are normalized: $\int (
R_0^2 + R_2^2 ) r^2 dr  =1$  and satisfy the system of differential equations (6) from  \onlinecite{Gelmont71,Rodina2010}, where the QD potential $V(r)$ instead of
the Coulomb one is taken.
Below we find $R_0$ and $R_2$ by numerical and variational methods.

\subsection{Numerical method}

To calculate numerically the energy spectrum and eigen wave functions of the hole in parabolic or Gaussian quantum dot numerically  we follow the approach described in Refs.~\onlinecite{Balderesci70,Balderesci73}. We diagonalize the hole Hamiltonian  matrix \cite{Gelmont71} calculated on non-orthogonal basis, consisting of Gaussian functions times the polynomials of the lowest power, which behave correctly at $r=0$:
 \begin{multline}\label{basis_sph}
 R_0=\sum\limits_{i=1}^{N_{\text{max}}=80}A_i\exp\left(-\alpha_i \tilde r^2\right),\\ ~R_2=\sum\limits_{i=1}^{N_{\text{max}}=80}B_i \tilde r \exp\left(-\alpha_i \tilde r^2\right).
 \end{multline}
Here $A_i$ and $B_i$ are coefficients which are to be found by the diagonalization  of the Hamiltonian matrix, $\alpha_i$ are chosen in the form of geometrical progression from  $10^{-6}$ to $10^3$.  The convergence of the calculation was controlled by modifying  the basis \eqref{basis_sph}: changing $\alpha_i$ and $N_{\text{max}}$. The calculation was believed to be converged if the basis modification did not change the result. Rather large $N_{\text{max}}$ as compared with \cite{Balderesci70} is necessary to obtain  reliable results in case of $\beta\rightarrow 0$, here $\beta=m_{lh}/m_{hh}$ is light to heavy hole effective mass ratio.
For the limiting case $\beta=1$, all $B_i=0$ with a good accuracy and the use of the basis \eqref{basis_sph} gives the same results as the use of the basis \eqref{harm}. For
 $\beta\rightarrow 0$ numerically calculated hole radial functions   satisfy  the  exact
 differential condition: \cite{Gelmont71} 
\be
\frac{dR_0}{dr}+\frac{dR_2}{dr} +\frac{3}{r}R_2 =0 \, .
\label{diffconn}
\ee
 with a  good accuracy.

\subsection{Variational method}

We chose the trial functions $R_0$ and $R_2$ for the arbitrary value of $\beta$ allowing them to satisfy the hole Hamiltonian \cite{Gelmont71}  in  two limits $\beta=1$ and $\beta=0$. If  $\beta=1$, the limiting case of the simple band dispersion is realized, and for the ground state the probe radial functions  should be chosen as  $R_2 =0$ and $R_0$ as given by \eqref{R0}.  The exact solution  for $\beta=0$  is not known, however, the functions  $R_0$ and $R_2$ must satisfy \eqref{diffconn}. Using these conditions and comparing the resulting  functions with  the numerically found solutions we arrived at:
\begin{multline}\label{VarR}
R_2(r)=\frac{C}{L^{3/2}} \cdot \frac{\alpha \tilde{r}^2}{2} \left[ \exp\left(-\frac{\alpha \tilde{r}^2}{2}\right) - \alpha_2 \exp\left(-\frac{\alpha \beta^{0.3} \tilde{r}^2}{2}\right) \right] \, ,\\
R_0(r)=\frac{C}{L^{3/2}}  \cdot \frac{3}{2} \left[ \exp\left(-\frac{\alpha \tilde{r}^2}{2}\right) + \alpha_0 \exp\left(-\frac{\alpha \beta^{0.3} \tilde{r}^2}{2}\right)\right]- \\ - R_2(r) \, ,
\end{multline}
where  $\alpha$, $\alpha_0$ and  $\alpha_2$  are the trial parameters and $C$ is the
normalization constant.   The oscillator length $L$ is defined as for the single band
with heavy hole effective mass $m_{ hh}$ and $\tilde{r}=r/L$. Note, that
taking $\alpha_0=\beta^{3/2}$ and $\alpha_2=\beta^2$ and using $\beta^{0.5}$ instead of
$\beta^{0.3}$ in  the second exponent in $R_0$ and $R_2$ we  arrive to the trial function
used in  Ref. \onlinecite{Rodina2010} for the parabolic confinement potential.

\subsection{Results: ground state energy and radial wave functions}

\begin{figure}[hptb]
\begin{center}\includegraphics[width=0.8\columnwidth]{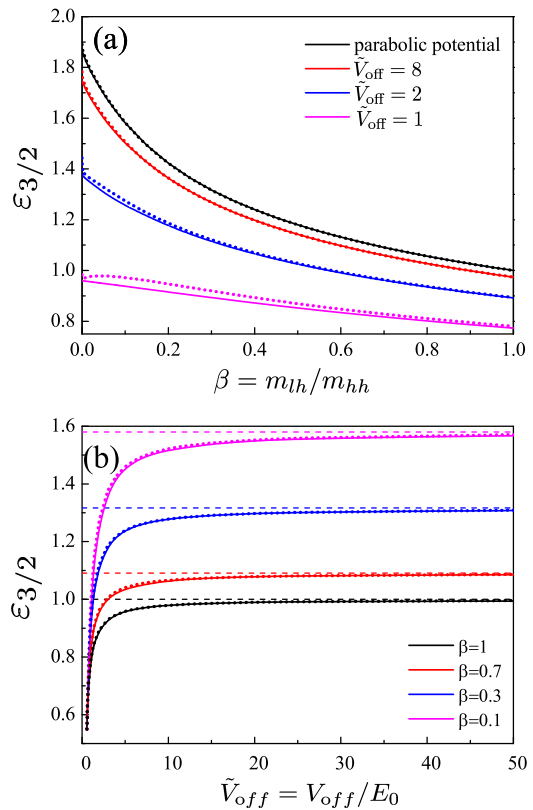}
\end{center}
\caption{ (Color online) Dimensionless ground state energy $\varepsilon_{3/2}$ of the hole as the function of $\beta$ for parabolic potential and Gaussian potential with $\tilde V_\text{off}=8;2;1$  (a) and as the function of $\tilde V_\text{off}$ for $\beta=0.1;0.3;0.7;1$ (b).  Solid lines correspond to numerical calculation, dotted lines, to variational calculation. Dashed lines on panel (b) indicate the energy in the limit of harmonic oscillator for respective values of $\beta$ . }\label{energy_fig}
\end{figure}

\begin{figure}[hptb]
\begin{center}\includegraphics[width=0.8\columnwidth]{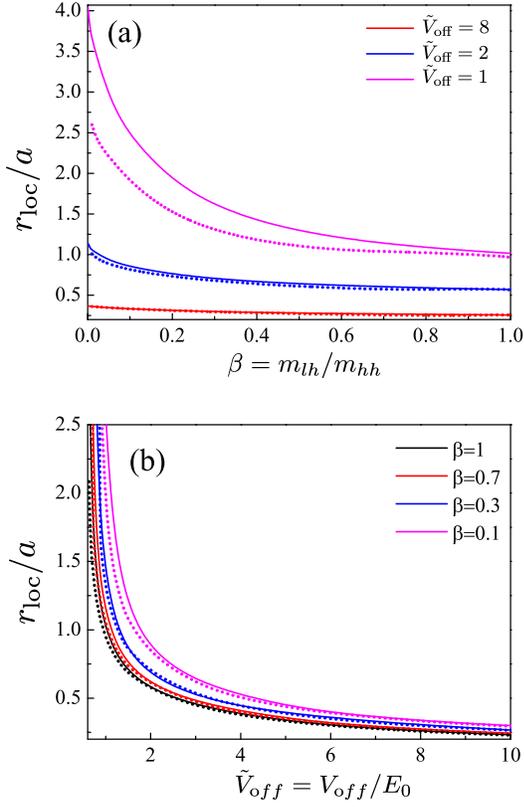}
\end{center}
\caption{ (Color online)
Ratio  of the localization radii and the characteristic dot radii,  $r_{\rm loc}/a$,
 the function of $\beta$ for parabolic potential and Gaussian potential with $\tilde V_\text{off}=8;2;1$ (a) and as the function
 of $\tilde V_\text{off}$ for $\beta=0.1;0.3;0.7;1$ (b).
  Solid lines correspond to numerical calculation, dotted lines, to variational calculation.  }\label{loc_fig}
\end{figure}

\begin{figure}[hptb]
\begin{center}
\includegraphics[width=0.7\columnwidth]{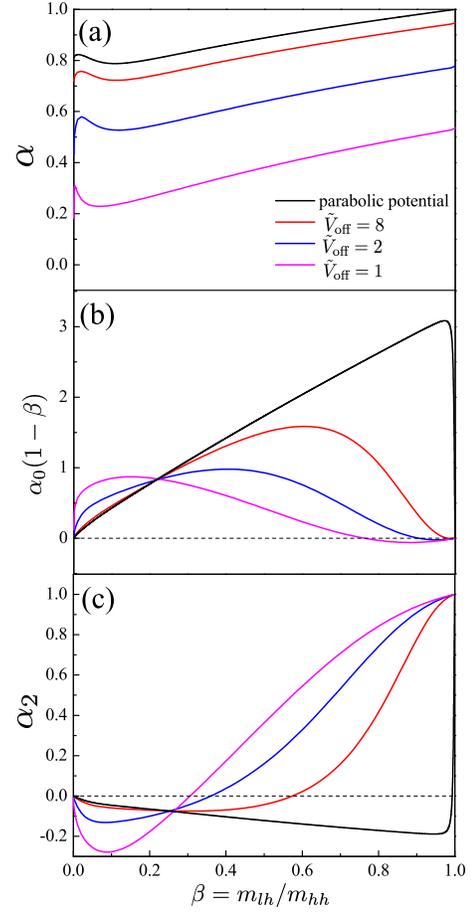}
\end{center}
\caption{\label{var} (Color online)  Dependencies of the variational parameters $\alpha$ (a), $\alpha_0 (1-\beta)$  (b) and $\alpha_2$ (c)  on $\beta$ for parabolic potential and  $\tilde V_\text{off}=8;2;1$ and dependencies of the variational parameter $\alpha$ on $\tilde V_\text{off}$ for $\beta=0.1;0.3;0.7;1$ (d).}
\end{figure}

The ground state energy, $E_{1S_{3/2}}(\beta)$, is expressed  in the units of $E_0$ (with $m^{*}=m_{\rm h}$) as follows:
\bea
E_{1S_{3/2}}(\beta)=
\frac{3}{2}\frac{\hbar^2}{m_{h}L^2} \epsilon_{3/2}(\beta) = E_0 \epsilon_{3/2}(\beta)\, .
\eea
The dimensionless function $\epsilon_{3/2}(\beta)$ calculated variationally and  numerically is shown in Fig. \ref{energy_fig}(a) for the parabolic potential and for Gaussian potential with  $\tilde V_\text{off}=8;2;1$.  Fig. \ref{energy_fig}(b)    shows $\epsilon_{3/2}$ as a function of $\tilde V_\text{off}$ for $\beta=0.1;0.3;0.7;1$. There is a good matching between two methods  demonstrating the high  accuracy of the variational method, which slightly decreases only for  very shallow dot potential (small $\tilde{V}_{\text{off}}$ or very small $\beta$). This fact allows us to validate our choice of the trial function in the form \eqref{VarR}.   The critical value of $\tilde{V}_{\text{off}}$ defining the appearance of the hole bound state increases with the decrease of $\beta$. Figures \ref{loc_fig} (a) and (b) show the dependencies  of the ratio  $r_{\rm loc}/a$ on  $\tilde{V}_\text{off}$ and $\beta$,  localization radius,  $r_{\rm loc}$, is defined as  $r_{\rm loc} =\sqrt{< r^2>}=\sqrt{\int_0^\infty ({R}_0^2 + {R}_2^2) r^4 d  r} $. Fig. \ref{var} shows  the dependencies of the variational parameters $\alpha$, $\alpha_0 (1-\beta)$ and $\alpha_2$ on $\beta$ for $\tilde V_\text{off}=8;2;1$.

\section{Anisotropic splitting of the hole ground state}\label{anis_splitting}

In this section we consider the hole states in the ellipsoidal QDs with $V(\bm r)$ given by Eq. (\ref{an}) and by Eq. (\ref{anGfull}). Additionally, we consider the effect of the internal crystal field  in wurtzite semiconductors, for example CdSe, in the framework of the quasi-cubic approximation. The respective addition to the Hamiltonian  is described by $\hat V^{\rm cr}=\Delta_{\rm cr}(5/8-j_z^2/2)$, where $\Delta_{\rm cr}$ is the energy splitting of the light hole and heavy hole valence  band edge  states in the bulk semiconductor.\cite{BirPikus}

The internal crystal field in wurtzite semiconductors  and the axial anisotropy of the confinement potential lifts the  degeneracy of the $1S_{3/2}$ ground hole states in the quantum dot. The four-fold degenerate hole state is split into two doublets with $|M|=3/2$ and $|M|=1/2$:
\be
E_{1S_{3/2},~M} = E_{1S_{3/2}}^{a} +  \frac{\Delta}{2} \left[\frac{5}{4}-M^2\right] \, , \\
\label{splitting}
\ee
where $\Delta= \Delta_{\rm int}+\Delta E^{\rm a}$, $\Delta_{\rm int}$ describes the effect of the internal crystal field, $\Delta E^{\rm a}=E_{1S_{3/2},~1/2}-E_{1S_{3/2},~3/2}$ and  $E_{1S_{3/2}}^{a}=(E_{1S_{3/2},~1/2}+E_{1S_{3/2},~3/2})/2$, describe the hole ground state splitting and the energy shift induced by the QD anisotropy. We calculate $\Delta_{\rm int}$, $\Delta E^{\rm a}$, and $E_{1S_{3/2}}^{a}$ numerically (see method description below) and determine the range of parameters  where the action of the crystal field and QD shape anisotropy can be considered as a perturbation. For these parameters the  splittings are calculated by the perturbation theory combined with the variationally found radial wave functions $R_0$ and $R_2$ of the spherical approximation.

\subsection{Numerical method}
To describe the hole states in  non-spherical QDs in general one has to use the Hamiltonian $\hat H_{\rm L}+V(\bm r)$ with allowance for $V(\bm r)$ to be fully asymmetric. In order to calculate the hole energy spectrum and eigen functions we diagonalize the Hamiltonian matrix, calculated on orthonormal basis of anisotropic  harmonic oscillator eigenfunctions:
\begin{multline}\label{basis}
\Psi_{n_x,n_y,n_z}(x,y,z)=\psi_{n_x}(x)\psi_{n_y}(y)\psi_{n_z}(z),\\ ~n_x,n_y,n_z=1...N,
\end{multline}
where
\begin{multline}\psi_{n_t}(t)=\frac{1}{\sqrt
{2^{n_t}n_t!}}\left(\frac{1}{\pi\tilde{l}_t^2}\right)^{1/4}\exp\left(-\frac{t^2}{2\tilde{l}_t^2}\right)H_{n_t}\left(\frac{t}{\tilde{l}_t}\right),\\
t=x,y,z
\end{multline}
are the eigenfunctions of the harmonic oscillator with oscillator lengths $\tilde{l}_x = [\hbar^2(\gamma_1+\gamma)/m_0\kappa_x]^{1/4}, ~\tilde{l}_y = [\hbar^2(\gamma_1+\gamma)/m_0\kappa_y]^{1/4},~\tilde{l}_z=[\hbar^2(\gamma_1-2\gamma)/m_0\kappa_z]^{1/4}$.
Such a basis corresponds to hole eigenfunctions, formed by Bloch states with momentum projection $j_z=\pm 3/2$ on the direction of the anisotropy axis. In order to check the convergence of the calculation the second basis, corresponding to the holes formed by Bloch states with the momentum projection $j_z=\pm 1/2$ is used.
Note that even in isotropic case, where $\kappa_x=\kappa_y=\kappa_z$, $\tilde{l}_{x,y}\neq\tilde{l}_z$ due to difference of hole effective masses along coordinate axes. The possible asymmetry of this basis may make it possible to account better for the QD potential geometry and increase the convergence rate of the calculation.

 \subsection{Results: comparison of numerical and perturbational calculations}
\subsubsection{Effect of the internal crystal field}

\begin{figure}[hptb]
\begin{center}
\includegraphics[width=0.8\columnwidth]{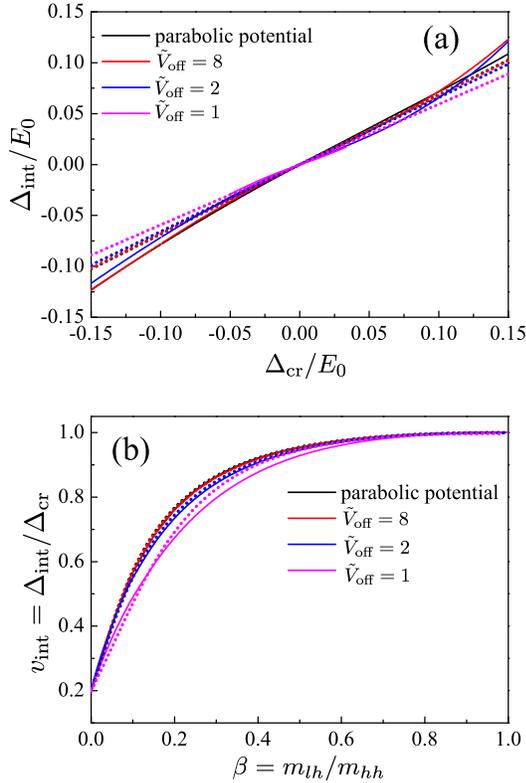}
\end{center}
\caption{(Color online) (a) The hole ground state splitting,  calculated  for $\beta=0.15$, as a function of parameter $\Delta_{\text{cr}}/E_0$ for   parabolic and Gaussian potential with $\tilde V_\text{off}=8;2;1$;    (b) The dimensionless function $v_{\text{int}}(\beta)$  as the function of the light-hole to heavy-hole effective mass ratio $\beta$. Solid lines - results of numerical calculations, dotted lines -  perturbation theory with the use of the trial functions.}\label{cryst}
\end{figure}

 The dependencies of  the hole ground state splitting  $\Delta_{\text{int}}$ on $\Delta_{\text{cr}}$, calculated numerically for the holes confined in parabolic and Gaussian potentials    are shown on Fig. \ref{cryst} (a) for $\beta=0.15$. It is clear that for small values of $\Delta_{\text{cr}}$ corresponding curves can be linearised, and the effect of the crystal field can be considered as perturbation \cite{EfrosPRB92,EfrosPRB96}:
\be \Delta_{\rm int}=\Delta_{\rm cr} v_{\rm int}= \Delta_{\rm cr}\int dr
r^2[R_0^2(r)-(3/5)R_2^2(r)]\, . \label{dint} \ee The function $v_{\rm int}$ depends   on
the ratio $\beta$ and, generally, may depend on the form of the QD potential. The
dependencies of   $v_{\rm int}(\beta)$  for  parabolic potential and Gaussian potential
with $\tilde V_\text{off}=8;2;1$ calculated numerically and obtained by using Eq.
\eqref{dint} with  the trial function  $R_0$ and $R_2$ in the form \eqref{VarR} are shown
in Fig. \ref{cryst} (b). A good agreement between two methods can be seen.
The function $v_{\rm int}(\beta)$  only slightly depends on the value of $\tilde
V_\text{off}$, but the $|M|=3/2$ states always correspond to the ground hole state
\cite{EfrosPRB92,EfrosPRB96}. The function  $v_{\rm int}$ increases from 0.2 for
$\beta=0$ to 1 for $\beta=1$. The value of $v_{\rm int}$  at $\beta=0$ is determined by
Eq. \eqref{diffconn} resulting in $\int R_2^2r^2dr=\int R_0^2 r^2 dr = 1/2$. For
$\beta=1$,  $v_{\rm int}=1$ is explained by the vanishing of $R_2$.

\subsubsection{Effect of the shape anisotropy}

\begin{figure}[hptb]
\begin{center}
\includegraphics[width=0.78\columnwidth]{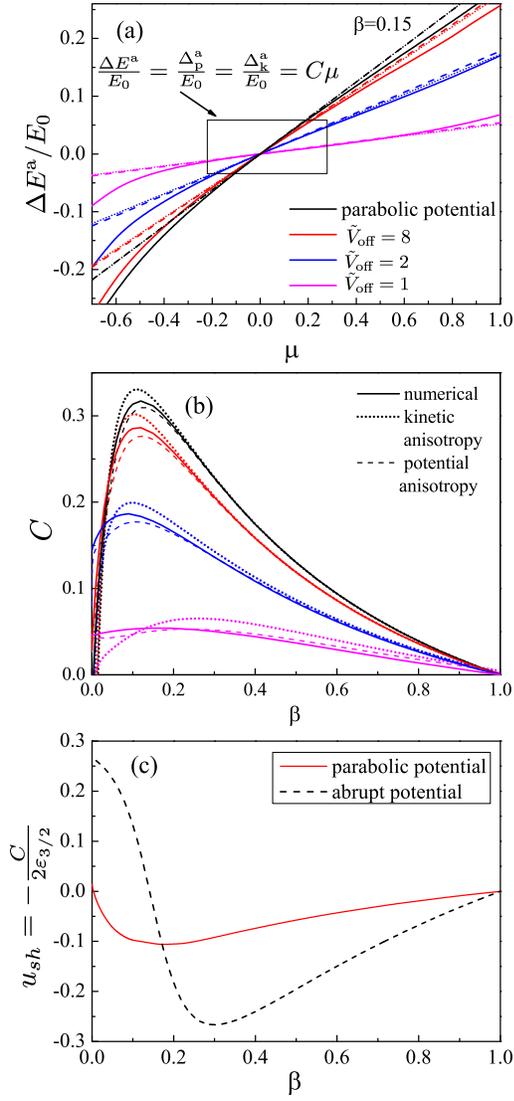}
\end{center}
\caption{ (Color online) (a)  The dimensionless splitting of the hole ground state, calculated numerically,
$\Delta E^{\rm a}/E_0$ (solid lines),  and by two perturbation theory methods:  $~\Delta_{\rm p}^{\rm a}/E_0$ (dotted lines),
 and $\Delta_{\rm k}^{\rm a}/E_0$ (dashed lines),  calculated with numerically obtained  functions $R_0(r)$ and $R_2(r)$ for parabolic and Gaussian QDs;
    (b) The coefficient of the linearisation, $C=\Delta E^{\rm a}/(E_0 \mu)$, calculated numerically (solid curves) and using the trial functions
    via perturbation in kinetic (dotted curves) and potential (dashed curves) energy;
    (c) The relative energy splitting,  $u_{sh}=-{\Delta E^{\rm a}}/{(2 \mu E_{3/2}}=-{C}/{2
\varepsilon_{3/2}}$,  calculated for the parabolic (solid curve) and abrupt infinite potential (dashed curve).}\label{splitting1}
\end{figure}

Figure \ref{splitting1} (a) shows the dependencies of the anisotropy induced  hole ground state splitting $\Delta E^{\rm a}$  calculated numerically for parabolic and Gaussian  potentials on anisotropy parameter $\mu$. Figure shows that in rather wide range of $\mu$ these dependencies can be approximated as linear with a good accuracy. In this case the splitting can be also found by the perturbation theory in two ways. One way is to consider as  the perturbation  the correction to the potential energy, $V_{\rm p}$ introduced in Eq. (\ref{an}) for parabolic and by Eq. (\ref{anG}) for the Gaussian potentials. For such a perturbation the hole energy splitting $\Delta E^{\rm a}=\Delta_{\rm p}^{\rm a}$ can be found as\cite{Rodina2010}
\be
\Delta_{\rm p}^{\rm a}=  \frac{4 \mu}{15}\frac{\hbar^2}{m_h L^4}
 \int dr r^4 R_0(r)R_2(r) \,  \label{delpot}
\ee

In the second way one can use the change of coordinates Eq. (\ref{axis}) in order to obtain the perturbation correction to the hole kinetic energy\cite{EfrosPRB93,misprint}:
\begin{multline}
 \hat H^{\rm a}_{\rm k}  = \frac{2\mu}{3} \frac{\hbar^2 } {2m_0} [ (\gamma_1 + \frac{5}{2}\gamma) (\hat k^2 - 3 \hat k_z^2)-\\ -2\gamma [ (\hat {\bm k}{\bm j})^2 - 3\{ (\hat {\bm k}{\bm j}){\hat k_z}j_z \} ] ] \, ,
\end{multline}
where $\{ ab \}=(ab+ba)/2$. Then the hole energy  splitting $\Delta E^{\rm a}=\Delta_{\rm k}^{\rm a}$ can be found as\cite{EfrosPRB93}
\bea
\Delta_{\rm k}^{\rm a}=
\frac{\mu \hbar^2}{3m_h}\left[I_1^{\rm a} -\frac{1}{5}I_2^{\rm a} +\frac{4}{5}I_3^{\rm a} -
\frac{1}{\beta}\left(I_1^{\rm a} -\frac{1}{5}I_2^{\rm a}  - \frac{4}{5}I_3^{\rm a} \right) \right] \, ,  \label{delpot1}
\eea
where
\begin{multline}
I_1^{\rm a}= \int r^2 dr \left[ \frac{dR_0(r)}{dr} \right]^2 \,, \\ ~I_2^{\rm a}= \int r^2 dr \left( \left[ \frac{dR_2(r)}{dr} \right]^2 + \frac{6R_2(r)^2}{r^2} \right) \, , \nonumber \\
I_3^{\rm a}= \int r^2 dr R_2(r) \left[ \frac{d^2R_0(r)}{dr^2}  - \frac{dR_0(r)}{rdr} \right] \, .
\end{multline}

Calculations using  radial wave functions $R_0$ and $R_2$ found  by the numerical method
results into $\Delta_{\rm p}^{\rm a}=\Delta_{\rm k}^{\rm a}=C\mu E_0$ with a good
accuracy for small $\mu$ shown in Fig. \ref{splitting1} (a) and  for all
values of light to heavy holes effective mass ratio $\beta$. Figure \ref{splitting1} (b)
shows the linear coefficients $C$ as functions of $\beta$. Dotted curves are calculated
numerically, solid and dashed curves correspond to $C=\Delta_{\rm k}^{\rm a}/\mu/E_0$ and
$C=\Delta_{\rm p}^{\rm a}/\mu/E_0$  calculated with the wave functions found via  the
variational procedure.  Figure shows, that the accuracy of the  variational method is
rather good. Moreover,  the kinetic energy and potential energy corrections found with
the variational wave functions  give the estimations of the  value of $C$ from above and
from below, respectively.

For the small values of $\mu$ linear corrections  to the  energy shift of the
central of level position   vanishes and $E_{1S_{3/2}}^{a}
=(E_{1S_{3/2},~1/2}+E_{1S_{3/2},~3/2})/2 \approx E_{1S_{3/2}}$. Numerical calculations
show that for complex valence band the following approximation
$E_{1S_{3/2}}^{a} \approx E_{1S_{3/2}} \left[1-\xi{\mu^2}/{9}  \right] $ is
valid, with factor $\xi$ being different from unity by no more than 10\%.
 Parameter
$\xi$ is the function of  $\beta$ and $V_{\text{off}}$. Figure
\ref{splitting1} (c) shows  the relative to the quantum size energy  splitting,
$u_{sh}=-\frac{\Delta E^{\rm a}}{2 \mu E_{3/2}}=-\frac{C}{2 \varepsilon_{3/2}}$,
introduced in \onlinecite{EfrosPRB93,EfrosPRB96} (the sign "-" is because of the
opposite definition of the sign of anisotropy parameter $\mu$) and calculated for
parabolic and abrupt potentials. One can see that there is the significant difference
between the dependencies caused by the different shape of the QD potential. The  shape
anisotropy at the abrupt potential in general induces much larger relative splitting than
that at the smooth one. Moreover, $u_{sh}$ changes sign for abrupt potential at
 $\beta\approx 0.15$, while remains always of the same sign in the
smooth potential.

\section{Hole effective $g$-factor}\label{magnetic_field}

In this section we consider the effect of  external magnetic field on the
holes states localized in quantum dots with the shape close to
spherical. For this purpose we follow the conventional
approach\cite{Luttinger,Roth,Gelmont73,EfrosPRB96,Rego97,Ivchenko,BirPikus,Kubisa11,Semina2015} and explore
the hole representation of  the Luttinger Hamiltonian with  the external magnetic field
${\bm B}$ included in spherical approximation.\cite{Gelmont73} In a weak magnetic field,
the top of the degenerated valence band is split according to the Zeeman term
\begin{equation}
\label{Zbulk}
\widehat{H}_Z=-2\mu_B\varkappa \left(\bm j\bm B\right).
\end{equation}
Here  $\mu_B$ -- is the Bohr magneton, $\varkappa$  is the Luttinger magnetic parameter,
and the lowest  valence hole state has projection $j_z= 3/2$ on the direction of
magnetic field for the semiconductors with $\varkappa>0$.

The effect of a weak external  magnetic field on the  holes confined  in some
potential of the spherical symmetry can be considered as the perturbation. The resulting
Zeeman splitting of the localized hole states is given by
\cite{Gelmont73,EfrosPRB96,EfrosCh3,Semina2015}
\begin{equation}
\label{Zbulk1}
\widehat{H}_{\rm eff}=-\mu_B g_{\rm h} \left(\bm J\bm B\right).
\end{equation}
Here $g_{\rm h}$ is the hole effective $g$-factor. For the  hole ground state
with $J=3/2$ its value can be determined via the radial wave functions $R_0$ and $R_2$
as:\cite{Gelmont73}
\begin{equation}\label{Gelmont}
g_{\rm h}   = 2\varkappa+\frac{8}{5}\gamma I_1^{\text{g}}+\frac{4}{5}\left[\gamma_1-2\left(\gamma+\varkappa\right)\right]I_2^{\text{g}} \, ,
\end{equation}
where
\begin{equation}\label{Gelmont1} I_1^{\text{g}}=\int\limits_{0}^{\infty}r^3 R_2(r)
\frac{d R_0(r)}{ dr} dr,~ I_2^{\text{g}}=\int\limits_{0}^{\infty}r^2 R_2^2(r) dr \, .
\end{equation}
The  integrals $I_1^{\text{g}}$ and $I_2^{\text{g}}$ describe the effect of
the light- and heavy-hole mixing induced by the confining potential. Their
values depend on the light- to heavy-hole effective mass ratio $\beta$ and do not depend
on the QD size.\cite{EfrosPRB96,EfrosCh3}  In the limit  $\beta=1$ the holes mixing
vanishes and $g_{\rm h}(\beta=1)=2\varkappa$. In the opposite case, $\beta \rightarrow
0$, the values of the mixing integrals can be found analytically as $I_1^{\text{g}}=-3/4$
and $I_2^{\text{g}}=1/2$ in any spherical symmetry potential using the differential
condition Eq. \eqref{diffconn}. This results in\cite{Gelmont73}
\begin{equation}\label{gbet0}
 g_{\rm h}(\beta=0)=\frac{6}{5}\varkappa + \frac{2}{5}\gamma_1-2\gamma \, .
\end{equation}
Using the relation $\varkappa=-2/3+5\gamma/3-\gamma_1/3,$\cite{EfrosCh3,DKK} we obtain
from Eq. \eqref{gbet0} that  $g_{\rm h}(\beta=0)\approx -0.8$ corresponding to the lowest
hole state with projection $M=-3/2$ on the magnetic field. Thus, in semiconductors with
small values of $\beta$ the mixing of the valence subbands may result in
different ordering of the Zeeman sublevels comparing with the free valence band edge
states.

\begin{figure}[hptb]
\begin{center}
\includegraphics[width=0.85\columnwidth]{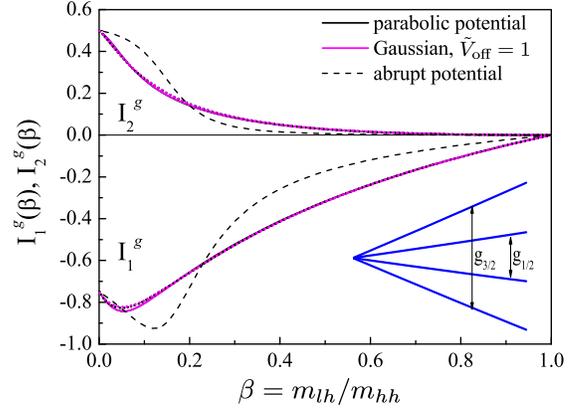}
\end{center}
\caption{(Color online)  Dimensionless integrals, $I_1^{\text{g}}$ and $I_2^{\text{g}}$, defining the hole ground state $g$-factor in spherical quantum dot, Eq. \eqref{Gelmont}, with  parabolic, Gaussian  with $\tilde V_\text{off}=1$ and abrupt infinite potential \cite{EfrosPRB96}   as a function of $\beta$.  Solid lines correspond to variational calculation, dotted lines to numerical calculation, dashed lines correspond to the abrupt infinite potential. }\label{fig_gfactor}
\end{figure}

We examine further the effect of the valence band mixing on the hole effective
$g$-factor in the QDs with different potential profiles. The dependencies of the mixing
integrals $I_1^{\text{g}}$ and $I_2^{\text{g}}$ on $\beta$, calculated variationally
(dotted lines) and numerically (solid lines) for  the parabolic (black lines) and
Gaussian  with $\tilde V_\text{off}=1$, smooth potentials and infinite abrupt
potential are shown on Fig. \ref{fig_gfactor}.  In fact, the
difference between values calculated variationally and numerically for smooth potentials is very small for any
mass ratio and can be hardly seen on  Figure, as well as the difference between values
calculated for the parabolic and Gaussian smooth potential. This demonstrates the
exceptional accuracy of the variational method. Moreover, $I_1^{\text{g}}$ and
$I_2^{\text{g}}$ are practically independent of $\tilde V_\text{off}$ (as long as a
confined level in Gaussian QD exists), while the hole wave functions are strongly
dependent on $\tilde V_\text{off}$ (see, for example, Fig. \ref{var} showing the resulting
trial parameters). As a result, the  $g$-factor for the hole in the parabolic and Gaussian
QDs can be estimated with a good accuracy using  the simple universal  approximation of the
$\beta$-dependence for mixing integrals:
\begin{multline}
I_1^{\text{g}}\approx e^{-5.145 \beta } (1-\beta ) \\ \left(-29.77 \beta ^5-37.97 \beta ^3+7.15 \beta ^2-7.77 \beta -0.75\right),\\
I_2^{\text{g}}\approx 0.5 e^{-7.35 \beta ^{1.127}} (1-\beta ) \left(14.23 \beta ^{2.58}+1\right).
\end{multline}  
In contrast, Fig. \ref{fig_gfactor}  shows the noticeable difference between
mixing integrals $I_1^{\text{g}}$ and $I_2^{\text{g}}$ calculated  for smooth potentials,
 and for  infinite abrupt potential. \cite{EfrosPRB96}  It means that the magneto-optical properties of QDs with smooth and abrupt potentials can be quite
 different.

 In the spherically-symmetric QDs, the hole ground state in zero magnetic field is fourfold degenerate with respect to the momentum projection $M$. Hence, in agreement with Eq. \ref{Zbulk1} the ground
  state   splits in weak magnetic fields into four equidistant sublevels (see the inset in Fig. \ref{fig_gfactor}).
   So, the two effective $g$-factors can be introduced in this case: $g_{1/2}=g_{\rm h} $
   for the splitting of the states with momentum projection $M \pm 1/2$ on the magnetic field direction
   and  $g_{3/2}=3 g_{\rm h}$ for the states with momentum projection $M=\pm 3/2$. The lowest hole state is always  $M=-3/2$ for semiconductors with $\beta \rightarrow 0$ and $M=+3/2$ ($M=-3/2$) for semiconductors with $\varkappa>0$ ($\varkappa<0$) and $\beta \rightarrow 1$.

In the axially symmetric QD the joint effect of the anisotropy and
magnetic field depends on the direction of magnetic field with respect to the direction
of the anisotropy axis $z$. In the case of small anisotropy, for ${\bm B} ||
z$ the hole effective $g$ factors for $M=\pm 3/2$ and $M=\pm 1/2$ states remain
the same: $g_{3/2}= 3g_{\rm h}$ and $g_{1/2}= g_{\rm h}$, respectively.
\cite{Semina2015}
 For the  ${\bm B} \perp z$ they become strongly anisotropic:
 the linear on ${\bm B}$ splitting of   $\pm 3/2$ states vanishes,  whereas the splitting of  $\pm 1/2$ states is described by the effective $g$ factor equal to $g_{1/2}^\perp=2g_{\rm
 h}$. Such consideration is valid while the magnetic field induced splitting is smaller than the zero field energy splitting $\Delta E^{\rm a}
 $. In addition, the  ${\bm B} \perp z$ mixes the hole states with $M=\pm 3/2$ and $M=\pm
 1/2$.\cite{EfrosPRB96,EfrosCh3}

 In the case of highly anisotropic QDs, i.e. in the limit of the quantum disk or quantum wire,
 it can be convenient to describe the hole states in the framework of  Luttinger spinors introduced in Ref. \onlinecite{Rego97}
  and classify the hole states by parity quantum number and $z$-component of the total angular momentum, which determine their
  splitting in the magnetic field. In these cases, the hole $g$-factors are substantially different from the values of $g_{3/2}$ and $g_{1/2}$, calculated above.
  The respective anisotropic $g$-factor values were calculated  in Ref. \onlinecite{Semina2015} for the model of the parabolic potential.

\section{Discussion}

In this paper  we presented the detailed study of the hole states in quantum dots with the 
smooth   potential  shape  of rotational symmetry, which  can be realized in II-VI
structures.  We developed  the variational approach with only three trial parameters,
which allowed us to calculate  with a good accuracy not only the hole energy but also its
$g$-factor and energy splitting of ground state caused by an anisotropy of the quantum
dot shape or the crystal field. These quantities occur to be strongly dependent on the wave
function and, therefore, very  sensitive to the choice of the trial function.
For example, we have found,  that the simplified  trial functions with the
only  trial parameter from Ref. \onlinecite{Rodina2010} predict the hole ground
state energy with a good accuracy, however do not allow to calculate the hole
$g$-factor and anisotropy--induced splittings.  The accuracy of the new trial
functions and the  developed variational method are verified by comparing the obtained results with
numerical calculations. The advantage of the variational method is to allow
one to have the hole envelope wave function in the simple analytical form that can be
rather easily  used for further modelling, e. g., of the multiexciton
states in the QDs.  At the same time, the developed numerical method allowed us to
calculate the whole energy spectrum including the excited states of the holes in the
smooth Gaussian--like potential.  To the best of our knowledge, neither variational method nor
numerical calculations (including the atomistic approaches) for  the hole states  in QDs
with such a  potential profile had not been reported before.

The general dependencies of the hole ground state characteristics  (energy and
localization radius) on QD potential depth and light to heavy hole effective mass ratio
are calculated for spherical QDs. The effect of  the QD anisotropy (potential shape or
internal crystal field) on the hole ground state is considered. The general dependencies
of the hole ground state splitting on potential depth and light to heavy hole effective
mass ratio are obtained. In addition, the Zeeman splitting of the hole ground state due
to the external magnetic field is studied. It is shown that the dependence of  the hole
effective $g$-factor  on the depth of the QD potential is negligible and its
dependence on $\beta$  for the QDs with the close to spherical shape can be
well approximated by the  universal analytical expressions. Moreover, the results
obtained in Ref. \onlinecite{Semina2015} for the hole effective $g$-factors in QDs
modelled by the ellipsoidal parabolic potential profiles of arbitrarily anisotropy can be
used for the case of the Gaussian-shaped QDs as well. Thus,  in the limit of
weak magnetic fields the effective hole $g$-factor is determined solely by the potential
profile type, but does not depend on it's size and barrier height. These results are in
line with known independence of the effective $g$-factor of the localization energy of
the hole bound to a deep or Coulomb--like acceptor
center\cite{Gelmont73,Malyshev,Malyshev2} and of the QD size.\cite{EfrosPRB96,EfrosCh3}  In contrast, the Zeeman splittings and the zero field splittings caused by the
QD shape anisotropy are quite different for the abrupt and the smooth confining
potentials. Such a  difference may include even different ordering
of the hole states both in zero and in external magnetic field.  Therefore, the combining of the smooth and abrupt potentials, e.g. by
variation of QD composition, opens new possibilities to designing the structures with the
needed properties of the hole states.

Let us discuss the applicability of the developed model to the II-VI QDs with gradually varying concentration. \cite{Peranio2000,Litvinov2008,Nasilowski15}  The potential profile in such dots indeed can be approximated by the Gaussian. However, the varying of concentration implies also     the spatial variation of the effective mass parameters. In II-VI QDs the variation of the effective mass parameters is not large and for the first approximation the mean values can be used with our model. The  energy corrections caused by the effective mass spatial variation are expected to be of the same order of magnitude as the corrections caused by the non-parabolicity of  the energy dispersion (terms $\propto \hat k^4$).\cite{Volkov99} Therefore,
they should be considered  in the framework of the Kane ${\bm k }{\bm p}$ model taking
into account the interaction between conduction and valence bands and that may be a
subject of a future study.

\section{acknowledgments}
The authors are thankful to T.V. Shubina and R.A. Suris for stimulating discussions.  The
work was  supported by Russian Science Foundation (Project number 14-22-00107).


\begin{thebibliography}{99}
\bibitem{note} Note, that the factor $(i)^l$ introduced because we use the definition of spherical harmonics $Y_{lm}$ as given in Ref. \onlinecite{Edmonds} while  in Ref. \onlinecite{Gelmont71} the spherical harmonics were defined according to Ref. \onlinecite{LL} .
\end{thebibliography}
\end{document}